\documentclass[10pt,twocolumn,superscriptaddress,amssymb,amsmath,aps,pra]{revtex4-1}
\usepackage{graphicx}

\newcommand*{\Scale}[2][4]{\scalebox{#1}{$#2$}}

\begin{document}

\title{Generalized noise terms for the quantized fluctuational electrodynamics}
\date{December 21, 2016}
\author{Mikko Partanen}
\affiliation{Engineered Nanosystems group, School of Science, Aalto University, P.O. Box 12200, 00076 Aalto, Finland}
\author{Teppo H\"ayrynen}
\affiliation{Engineered Nanosystems group, School of Science, Aalto University, P.O. Box 12200, 00076 Aalto, Finland}
\affiliation{DTU Fotonik, Department of Photonics Engineering, Technical University of
Denmark, \O rsteds Plads, Building 343, DK-2800 Kongens Lyngby, Denmark}
\author{Jukka Tulkki}
\affiliation{Engineered Nanosystems group, School of Science, Aalto University, P.O. Box 12200, 00076 Aalto, Finland}
\author{Jani Oksanen}
\affiliation{Engineered Nanosystems group, School of Science, Aalto University, P.O. Box 12200, 00076 Aalto, Finland}

\begin{abstract}
The quantization of optical fields in vacuum
has been known for decades,
but extending the field quantization to lossy and dispersive media
in nonequilibrium conditions has
proven to be complicated due to the position-dependent
electric and magnetic responses of the media.
In fact, consistent position-dependent quantum models for the photon number
in resonant structures have only been formulated very recently and only for dielectric media.
Here we present a general position-dependent quantized fluctuational
electrodynamics (QFED) formalism that extends the consistent
field quantization to describe the
photon number also in the presence of magnetic field-matter interactions.
It is shown that the magnetic fluctuations provide an additional
degree of freedom in media where the magnetic
coupling to the field is prominent. Therefore, the field quantization
requires an additional independent noise operator that is
commuting with the conventional bosonic noise operator describing the
polarization current fluctuations in dielectric media.
In addition to allowing the detailed description of field fluctuations,
our methods provide practical tools for modeling optical
energy transfer and the formation of thermal balance
in general dielectric and magnetic nanodevices.
We use QFED to investigate the magnetic properties of microcavity
systems to demonstrate an example geometry in which
it is possible to probe fields arising from the electric
and magnetic source terms.
We show that, as a consequence of
the magnetic Purcell effect, the tuning of
the position of an emitter layer placed inside a vacuum cavity
can make the emissivity of a magnetic emitter to
exceed the emissivity of a corresponding
electric emitter.
\end{abstract}

\maketitle

\section{Introduction}

Better understanding of optical phenomena and nanoscale energy transfer has
enabled advances in optical technologies, e.g., in nanoplasmonics
\cite{Sorger2011,Oulton2009,Huang2013,Sadi2013},
near-field microscopy \cite{Taubner2006,Hillenbrand2002},
thin-film light-emitting diodes \cite{Nakamura2013,Heikkila2013},
photonic crystals \cite{Russell2003,Akahane2003}, and metamaterials \cite{Tanaka2010,Mattiucci2013}.
These advances are strongly
influenced by the availability of simple and transparent theoretical tools
that allow in-depth understanding of the pertinent phenomena
in sufficiently simple form.
Formulating a simple and sufficiently detailed description
of the quantum aspects of energy transfer in lossy resonant structures, however,
has been particularly challenging due to several phenomena, such as
wave-particle dualism, intertwined electric and magnetic
fields, and field-matter interactions \cite{Shadbolt2014,Altmann2015},
affecting the energy transfer.

We have recently introduced quantized fluctuational electrodynamics (QFED) formalism
\cite{Partanen2014a,Partanen2014c,Partanen2015a,Partanen2014b} for the description of
field-matter interactions and the formation of thermal balance in nonequilibrium conditions
in dielectric media.
Using the QFED approach it has finally become possible to formulate the canonical
commutation relations preserving ladder and photon-number operators
for the total electromagnetic field also in resonant structures
\cite{Partanen2014a,Partanen2014c,Partanen2015a},
thus resolving the previously reported anomalies in their
commutation relations
\cite{Raymer2013,Barnett1996,Ueda1994,Aiello2000,Stefano2000}.
Here we present a more general QFED formalism that
also accounts for the interactions arising due to magnetic effects.
The need to formulate the unified theory emerges from the
present observation that the single electric
noise operator picture is insufficient
to correctly describe the field-matter interactions
and the photon annihilation operator of a single polarization
in the general case.
Instead, the use of two commuting bosonic noise source operators is needed
to formulate the general model. By using two separate operators
it becomes straightforward to develop the QFED model starting from
the macroscopic Maxwell's equations
and the polarization and magnetization related material responses.
Overall, however, the requirement of using two commuting noise operators
to allow consistent field quantization presents
a significant change in the conventional description of the photon
annihilation operator in interfering structures.

After deriving the generalized QFED formalism, we apply it to show that,
due to the magnetic Purcell effect,
we can tune the position of the emitter in a vacuum cavity such
that the emissivity of a magnetic emitter significantly exceeds
the emissivity of a corresponding electric emitter. This is
a consequence of different position dependences of the electric
and magnetic local densities of states (LDOSs).
The differences in emissivities are expected to be experimentally observable
by detecting the output radiation with a photodetector or an antenna.

This manuscript is organized as follows: The theory
is presented in Sec.~\ref{sec:theory}. It covers the introduction
of the quantized equations, representation of the Green's functions and
the related noise source operators. We also briefly review the theory of field fluctuations,
photon numbers, Poynting vector, and absorption and emission operators
while presenting the new generalized forms of the densities of states.
In Sec.~\ref{sec:results},
we investigate the physical implications of the concepts by applying
the methods to study the emissivity of electric and magnetic emitters
placed inside the vacuum cavity and demonstrate the different
characteristics of the two fundamentally different source terms
of our theory. Finally, conclusions are drawn in Sec.~\ref{sec:conclusions}.

\section{\label{sec:theory}Field quantization}

\subsection{\label{sec:Maxwell}Quantized equations}

Maxwell's equations relate the electric field strength $\mathbf{E}$,
the magnetic field strength $\mathbf{H}$,
the electric flux density $\mathbf{D}$, and
the magnetic flux density $\mathbf{B}$ to the free electric
charge $\rho_\mathrm{f}$ and current $\mathbf{J}_\mathrm{f}$
densities \cite{Jackson1999}. In the frequency domain, the equations
for positive frequencies read
\begin{align}
 \nabla\cdot\mathbf{D} &=\rho_\mathrm{f},\label{eq:maxwell1}\\
 \nabla\cdot\mathbf{B} &=0,\label{eq:maxwell2}\\
 \nabla\times\mathbf{E} &=i\omega\mathbf{B}=i\omega\mu_0(\mu\mathbf{H}+\delta\mathbf{M}),\label{eq:maxwell3}\\
 \nabla\times\mathbf{H} &=\mathbf{J}_\mathrm{f}-i\omega\mathbf{D}=\mathbf{J}_\mathrm{f}-i\omega(\varepsilon_0\varepsilon\mathbf{E}+\delta\mathbf{P}),\label{eq:maxwell4}
\end{align}
where we have related the fields and field densities in Eqs.~\eqref{eq:maxwell3} and \eqref{eq:maxwell4}
using the constitutive relations $\mathbf{D}=\varepsilon_0\varepsilon\mathbf{E}+\delta\mathbf{P}$
and $\mathbf{B}=\mu_0(\mu\mathbf{H}+\delta\mathbf{M})$,
where $\varepsilon_0$ and $\mu_0$
are the permittivity and permeability of vacuum,
$\varepsilon=\varepsilon_\mathrm{r}+i\varepsilon_\mathrm{i}$
and $\mu=\mu_\mathrm{r}+i\mu_\mathrm{i}$
are the relative permittivity and permeability of the medium
with real and imaginary
parts denoted by subscripts $\mathrm{r}$ and $\mathrm{i}$,
and the polarization and magnetization fields
$\delta\mathbf{P}$ and $\delta\mathbf{M}$ denote the polarization
and magnetization that are not linearly proportional to the respective field
strengths \cite{Sipe1987}. In the context of the present work,
$\delta\mathbf{P}$ and $\delta\mathbf{M}$ describe
small noise related parts in the linear polarization and 
magnetization fields as customary in the classical
fluctuational electrodynamics \cite{Joulain2005}.

From the Maxwell's equations in Eqs.~\eqref{eq:maxwell1}--\eqref{eq:maxwell4}
it follows that the electric and the magnetic fields obey the well-known equations
\begin{align}
 \nabla\times\Big(\frac{\nabla\times\mathbf{E}}{\mu_0\mu}\Big)-\omega^2\varepsilon_0\varepsilon\mathbf{E} &=i\omega\mathbf{J}_\mathrm{e}
 -\nabla\times\Big(\frac{\mathbf{J}_\mathrm{m}}{\mu_0\mu}\Big),\label{eq:HelmholtzE1}\\
 \nabla\times\Big(\frac{\nabla\times\mathbf{H}}{\varepsilon_0\varepsilon}\Big)-\omega^2\mu_0\mu\mathbf{H} &=i\omega\mathbf{J}_\mathrm{m}
 +\nabla\times\Big(\frac{\mathbf{J}_\mathrm{e}}{\varepsilon_0\varepsilon}\Big)\label{eq:HelmholtzB1},
\end{align}
where the terms $\mathbf{J}_\mathrm{e}=\mathbf{J}_\mathrm{f}-i\omega\delta\mathbf{P}$
and $\mathbf{J}_\mathrm{m}=-i\omega\mu_0\delta\mathbf{M}$ represent
the polarization and magnetization noise currents that act as field
sources also in the classical fluctuational electrodynamics
\cite{Narayanaswamy2014,Polimeridis2015}.
The electric term $\mathbf{J}_\mathrm{e}$ includes contributions from both the electric currents due to free charges (which amount
to zero for insulating dielectrics) as well as polarization terms associated with
dipole currents and thermal dipole fluctuations. For the magnetic term $\mathbf{J}_\mathrm{m}$,
the only contribution arises from the magnetic dipoles.

For simplicity, we limit the present analysis to the case of normal incidence
in a structure where the material parameters only depend on the position
coordinate $x$ and
formulate Eqs.~\eqref{eq:HelmholtzE1} and \eqref{eq:HelmholtzB1}
as a single polarization scalar problem
where the electric and magnetic fields
are parallel to the $y$ and $z$ axes, respectively.
In the QFED framework, the components of the classical fields and currents
in Eqs.~\eqref{eq:maxwell1}--\eqref{eq:HelmholtzB1}
are replaced by corresponding quantum field operators
$\hat E^+(x,\omega)$ and $\hat H^+(x,\omega)$ and noise current operators
$\hat J_\mathrm{e}^+(x,\omega)$ and
$\hat J_\mathrm{m}^+(x,\omega)$ to account for the quantum features
of the field and the noise statistics.
From Eqs.~\eqref{eq:HelmholtzE1} and \eqref{eq:HelmholtzB1}
it follows that, in our case of normal incidence,
the noise current operators $\hat J_\mathrm{e}^+(x,\omega)$ and
$\hat J_\mathrm{m}^+(x,\omega)$ describe noise current components
that are parallel to the electric and magnetic fields,
respectively. In the scalar form, the equations in
Eqs.~\eqref{eq:HelmholtzE1} and \eqref{eq:HelmholtzB1} then simplify to
\begin{align}
 &\frac{\partial}{\partial x}\Big(\frac{\partial \hat E^+(x,\omega)}{\mu_0\mu(x,\omega)\partial x}\Big)+\omega^2\varepsilon_0\varepsilon(x,\omega)\hat E^+(x,\omega)\nonumber\\
 &=-i\omega\hat J_\mathrm{e}^+(x,\omega)-\frac{\partial}{\partial x}\Big(\frac{\hat J_\mathrm{m}^+(x,\omega)}{\mu_0\mu(x,\omega)}\Big),\label{eq:HelmholtzE2}\\
 &\frac{\partial}{\partial x}\Big(\frac{\partial \hat H^+(x,\omega)}{\varepsilon_0\varepsilon(x,\omega)\partial x}\Big)+\omega^2\mu_0\mu(x,\omega)\hat H^+(x,\omega)\nonumber\\
 &=-i\omega\hat J_\mathrm{m}^+(x,\omega)-\frac{\partial}{\partial x}\Big(\frac{\hat J_\mathrm{e}^+(x,\omega)}{\varepsilon_0\varepsilon(x,\omega)}\Big).\label{eq:HelmholtzB2}
\end{align}
Note that these equations are not independent as
either equation allows fully solving the system, leaving the calculation of
the remaining fields a simple task when applying the appropriate Maxwell's equation.
In the following, we will mainly use Eq.~\eqref{eq:HelmholtzE2} as the starting point for further analysis.

\subsection{Green's functions}

In order to write the solution of Eq.~\eqref{eq:HelmholtzE2}
in a general form, we first define the electric Green's function
$G_\mathrm{ee}(x,\omega,x')$ that satisfies
\begin{equation}
 \frac{\partial}{\partial x}\Big(\frac{\partial G_\mathrm{ee}(x,\omega,x')}{\mu(x,\omega)\partial x}\Big)+k_0^2\varepsilon(x,\omega)G_\mathrm{ee}(x,\omega,x') =-\delta(x-x'),
 \label{eq:Greenee}
\end{equation}
where $k_0=\omega/c$ is the wavenumber in vacuum with the vacuum velocity of light $c$.
In terms of the electric Green's function, the solution of Eq.~\eqref{eq:HelmholtzE2} is written as
%
\begin{equation}
 \Scale[0.96]{
 \begin{array}{l}
 \hat{E}^+(x,\omega)\\[6pt]
 \displaystyle= \mu_0\int_{-\infty}^\infty\hspace{-0.2cm} G_\mathrm{ee}(x,\omega,x')\Big[i\omega\hat J_\mathrm{e}^+(x',\omega)
 +\frac{\partial}{\partial x'}\Big(\frac{\hat J_\mathrm{m}^+(x',\omega)}{\mu_0\mu(x',\omega)}\Big)\Big]dx'\\[12pt]
 \displaystyle=i\omega\mu_0\int_{-\infty}^\infty G_\mathrm{ee}(x,\omega,x')\hat J_\mathrm{e}^+(x',\omega)dx'\\[12pt]
 \displaystyle\hspace{0.5cm}+k_0\int_{-\infty}^\infty G_\mathrm{em}(x,\omega,x')\hat J_\mathrm{m}^+(x',\omega)dx',
 \end{array}
 }
 \label{eq:efield}
\end{equation}
where, in the case of the second term, we have applied integration by parts
with the boundary condition that the Green's functions go to zero at infinities as
they are exponentially decaying in lossy media and lossless media can be described
in the limit of small losses. We have also defined the
exchange Green's function $G_\mathrm{em}(x,\omega,x')$ as
\begin{equation}
 G_\mathrm{em}(x,\omega,x') =-\frac{\partial G_\mathrm{ee}(x,\omega,x')}{k_0\mu(x',\omega)\partial x'}.\label{eq:greenem}
\end{equation}

Solving for the magnetic field in the Maxwell's equation in Eq.~\eqref{eq:maxwell3}
and substituting the electric field operator in terms of the Green's functions in Eq.~\eqref{eq:efield} gives
\begin{equation}
 \Scale[0.96]{
 \begin{array}{l}
 \hat{H}^+(x,\omega)\\[6pt]
 \displaystyle=\frac{1}{i\omega\mu_0\mu(x,\omega)}\Big(\hat J_\mathrm{m}^+(x,\omega)+\frac{\partial \hat E^+(x,\omega)}{\partial x}\Big)\\[12pt]
 \displaystyle=\frac{1}{i\omega\mu_0\mu(x,\omega)}\Big(\hat J_\mathrm{m}^+(x,\omega)+i\omega\mu_0\int_{-\infty}^\infty\frac{\partial G_\mathrm{ee}(x,\omega,x')}{\partial x}\\[12pt]
 \displaystyle\hspace{0.5cm}\times\hat J_\mathrm{e}^+(x',\omega)dx'+k_0\int_{-\infty}^\infty\frac{\partial G_\mathrm{em}(x,\omega,x')}{\partial x}\hat J_\mathrm{m}^+(x',\omega)dx'\Big)\\[12pt]
 \displaystyle=k_0\int_{-\infty}^\infty\frac{\partial G_\mathrm{ee}(x,\omega,x')}{k_0\mu(x,\omega)\partial x}\hat J_\mathrm{e}^+(x',\omega)dx'\\[12pt]
 \displaystyle\hspace{0.5cm}-\frac{ik_0^2}{\omega\mu_0}\int_{-\infty}^\infty\Big[\frac{\partial G_\mathrm{em}(x,\omega,x')}{k_0\mu(x,\omega)\partial x}+\frac{\delta(x-x')}{k_0^2\mu(x,\omega)}\Big]\hat J_\mathrm{m}^+(x',\omega)dx'\\[12pt]
 \displaystyle=k_0\int_{-\infty}^\infty G_\mathrm{me}(x,\omega,x')\hat J_\mathrm{e}^+(x',\omega)dx'\\
 \displaystyle\hspace{0.5cm}+i\omega\varepsilon_0\int_{-\infty}^\infty G_\mathrm{mm}(x,\omega,x')\hat J_\mathrm{m}^+(x',\omega)dx',
 \end{array}
 }
 \label{eq:hfield}
\end{equation}
where we have defined the magnetic Green's function $G_\mathrm{mm}(x,\omega,x')$
and the exchange Green's function $G_\mathrm{me}(x,\omega,x')$ as
\begin{equation}
 G_\mathrm{me}(x,\omega,x')=\frac{\partial G_\mathrm{ee}(x,\omega,x')}{k_0\mu(x,\omega)\partial x},
 \label{eq:greenme}
\end{equation}
\begin{equation}
 G_\mathrm{mm}(x,\omega,x')=-\frac{\partial G_\mathrm{em}(x,\omega,x')}{k_0\mu(x,\omega)\partial x}-\frac{\delta(x-x')}{k_0^2\mu(x,\omega)}.
 \label{eq:greenmm}
\end{equation}
By using Eqs.~\eqref{eq:greenem} and \eqref{eq:greenmm}, one also obtains an
expression of the magnetic Green's function $G_\mathrm{mm}(x,\omega,x')$ directly in terms of the
electric Green's function $G_\mathrm{ee}(x,\omega,x')$ as
\begin{equation}
 G_\mathrm{mm}(x,\omega,x') =\frac{\partial^2G_\mathrm{ee}(x,\omega,x')}{k_0^2\mu(x,\omega)\mu(x',\omega)\partial x\partial x'}
 -\frac{\delta(x-x')}{k_0^2\mu(x,\omega)}.
 \label{eq:greenmm2}
\end{equation}
In Eq.~\eqref{eq:greenmm2}, the first term has a discontinuity at $x=x'$ due to the discontinuity
of the second order derivative of $G_\mathrm{ee}(x,\omega,x')$. However,
this discontinuity is completely balanced by the second term rendering
$G_\mathrm{mm}(x,\omega,x')$ continuous everywhere.

The electric and magnetic Green's functions obey the general reciprocity relations
$G_\mathrm{ee}(x,\omega,x')=G_\mathrm{ee}(x',\omega,x)$ and
$G_\mathrm{mm}(x,\omega,x')=G_\mathrm{mm}(x',\omega,x)$ \cite{Narayanaswamy2014}.
The reciprocity relation for the exchange Green's functions
$G_\mathrm{me}(x,\omega,x')=-G_\mathrm{em}(x',\omega,x)$,
follows from the definitions in Eqs.~\eqref{eq:greenem} and \eqref{eq:greenme}
and the reciprocity relations of $G_\mathrm{ee}(x,\omega,x')$
and $G_\mathrm{mm}(x,\omega,x')$.

The Green's functions depend on the problem geometry via the material permittivity
and permeability and they are continuous at material interfaces which
follows from the continuity of the electric and magnetic fields
$\hat E^+(x,\omega)$ and $\hat H^+(x,\omega)$.
For example, in a homogeneous space the electric and magnetic Green's functions
$G_\mathrm{ee}(x,\omega,x')$ and $G_\mathrm{mm}(x,\omega,x')$ are
\begin{align}
 G_\mathrm{ee}(x,\omega,x') &=\mu(\omega)\frac{ie^{ik(\omega)|x-x'|}}{2k(\omega)},\\
 G_\mathrm{mm}(x,\omega,x') &=\varepsilon(\omega)\frac{ie^{ik(\omega)|x-x'|}}{2k(\omega)},
\end{align}
where $k(\omega)=k_0n(\omega)$ is the wavenumber in the medium.
A simple method to calculate the Green's functions in
more general stratified media is described in Appendix \ref{apx:green}.

\subsection{\label{sec:sources}Noise operators}

In order to determine the forms of the noise current operators
$\hat J_\mathrm{e}^+(x,\omega)$ and $\hat J_\mathrm{m}^+(x,\omega)$,
we require that the resulting electric and magnetic field operators,
related to the noise source operators by Eq.~\eqref{eq:HelmholtzE2}
and \eqref{eq:HelmholtzB2}, obey
the well-known canonical commutation relations, i.e.,
$[\hat{A}(x,t),\hat{E}(x',t)]=-i\hbar/(\varepsilon_0S)\delta(x-x')$
\cite{Barnett1995,Barnett1996,Matloob1995}.
For purely dielectric media, studied in
Refs.~\cite{Partanen2014a} and \cite{Matloob1995},
the electric noise current operator is directly proportional to
a bosonic annihilation operator $\hat f(x,\omega)$ satisfying the
canonical commutation relation
$[\hat f(x,\omega),\hat f^\dag(x',\omega')]=\delta(x-x')\delta(\omega-\omega')$
through $\hat J^+(x,\omega)=j_\mathrm{0}(x,\omega)\hat f(x,\omega)$ where
$j_\mathrm{0}(x,\omega)$ is a normalization factor given by
$j_\mathrm{0}(x,\omega)=\sqrt{4\pi\hbar\omega^2\varepsilon_0\mathrm{Im}[n(x,\omega)^2]/S}$,
in which $S$ is the area of quantization in the $y$-$z$ plane and
$\hbar$ is the reduced Planck constant.
The operator $\hat f(x,\omega)$ gives the local source field
number operator
$\hat\eta(x,\omega)=\int\hat f^\dag(x,\omega)\hat f(x',\omega')dx'd\omega'$
whose expectation value
$\langle\hat\eta(x,\omega)\rangle$ is given by the Bose-Einstein distribution as
$\langle\hat\eta(x,\omega)\rangle=1/[e^{\hbar\omega/[k_\mathrm{B}T(x)]}-1]$,
in which $T(x)$ is the possibly position-dependent temperature profile of the medium \cite{Partanen2014a}.
In the formalism for dielectrics,
the derivation of the coefficient $j_0(x,\omega)$
assumes the relation $\varepsilon(x,\omega)=n(x,\omega)^2$ \cite{Matloob1995},
which is not satisfied in the case of magnetic media.
It also follows that the canonical commutation relations
of fields would not be satisfied in the case of magnetic media if we just
neglected the magnetic noise current operator and assumed the same form
for the electric noise current operator as that in purely
dielectric media.

Thus, to preserve the commutation relations and to find
the correct form of the current operators, we allow
an additional degree of freedom to conform with the
addition of the magnetic noise current operator.
The simplest possible current operator form using two independent bosonic source field operators
$\hat f_\mathrm{e}(x,\omega)$ and $\hat f_\mathrm{m}(x,\omega)$
is $\hat J^+_\mathrm{e}(x,\omega)=j_\mathrm{0,e}(x,\omega)\hat f_\mathrm{e}(x,\omega)$ and
$\hat J_\mathrm{m}(x,\omega)=j_\mathrm{0,m}(x,\omega)\hat f_\mathrm{m}(x,\omega)$,
where  $j_\mathrm{0,e}(x,\omega)$ and $j_\mathrm{0,m}(x,\omega)$
are normalization factors.
The above forms can also be partly motivated
by the fact that the electric and magnetic
current operators $\hat J_\mathrm{e}^+(x,\omega)$ and
$\hat J_\mathrm{m}^+(x,\omega)$ describe currents in
different directions.
The bosonic source field operators $\hat f_\mathrm{e}(x,\omega)$ and
$\hat f_\mathrm{m}(x,\omega)$ are assumed to obey the same canonical commutation relation
$[\hat f_j(x,\omega),\hat f_k^\dag(x',\omega')]=\delta_{jk}\delta(x-x')\delta(\omega-\omega')$,
where $j,k\in\{\mathrm{e,m}\}$, as above. Similarly, they also define two separate
local source field number operators 
$\hat\eta_\mathrm{e}(x,\omega)$ and $\hat\eta_\mathrm{m}(x,\omega)$.
In the case of a thermal source field described by the Bose-Einstein distribution,
the expectation values $\langle\hat\eta_\mathrm{e}(x,\omega)\rangle$ and
$\langle\hat\eta_\mathrm{m}(x,\omega)\rangle$ are additionally equal and denoted by
$\langle\hat\eta(x,\omega)\rangle$.
The normalization factors $j_\mathrm{0,e}(x,\omega)$ and $j_\mathrm{0,m}(x,\omega)$
can be determined apart from the possible phase factors by
requiring that the vector potential and electric field operators obey
the canonical commutation relation
$[\hat{A}(x,t),\hat{E}(x',t)]=-i\hbar/(\varepsilon_0S)\delta(x-x')$
\cite{Barnett1995,Barnett1996,Matloob1995}.
As a result from the calculation presented in Appendix \ref{apx:commutation}, we obtain
$j_\mathrm{0,e}(x,\omega)=\sqrt{4\pi\hbar\omega^2\varepsilon_0\varepsilon_\mathrm{i}(x,\omega)/S}$
and
$j_\mathrm{0,m}(x,\omega)=\sqrt{4\pi\hbar\omega^2\mu_0\mu_\mathrm{i}(x,\omega)/S}$.
This essentially proves that neither of the two noise source
operators $\hat f_\mathrm{e}(x,\omega)$ and $\hat f_\mathrm{m}(x,\omega)$ can be neglected.

\subsection{Field fluctuations, photon numbers, and densities of states}

In the case of purely dielectric media, the formulas for the field
fluctuations, photon numbers, Poynting vector, and local densities of states
as described in QFED are
presented in Refs.~\citenum{Partanen2014a} and \citenum{Partanen2014c}.
In the present case, the general form of the equations stays the same
but the densities of states are substantially modified.
For completeness, we review below the general formulae while
presenting the new generalized nonlocal and interference densities of states.

The spectral components of the time domain
electric and magnetic field fluctuations and the energy density
$\langle\hat u(x,t)\rangle_\omega=\frac{1}{2}|\varepsilon_0\varepsilon(x,\omega)|\langle\hat E(x,t)^2\rangle_\omega+\frac{1}{2}|\mu_0\mu(x,\omega)|\langle\hat H(x,t)^2\rangle_\omega$
for a single polarization and angular frequency $\omega$
are written in terms of the photon-number expectation values as \cite{Partanen2014c}
\begin{align}
 \langle\hat E(x,t)^2\rangle_\omega & =\frac{\hbar\omega}{\varepsilon_0}\rho_\mathrm{e}(x,\omega)\Big(\langle\hat n_\mathrm{e}(x,\omega)\rangle+\frac{1}{2}\Big)\label{eq:efluct},\\[8pt]
 \langle\hat H(x,t)^2\rangle_\omega & =\frac{\hbar\omega}{\mu_0}\rho_\mathrm{m}(x,\omega)\Big(\langle\hat n_\mathrm{m}(x,\omega)\rangle+\frac{1}{2}\Big)\label{eq:bfluct},\\[8pt]
 \langle\hat u(x,t)\rangle_\omega & = \hbar\omega\rho_\mathrm{tot}(x,\omega)\Big(\langle\hat n_\mathrm{tot}(x,\omega)\rangle+\frac{1}{2}\Big)\label{eq:edensity}.
\end{align}
The photon-number expectation values $\langle\hat n_\mathrm{j}(x,\omega)\rangle$,
$j\in\{\mathrm{e,m,tot}\}$, in Eqs.~\eqref{eq:efluct}--\eqref{eq:edensity} are given by
\begin{equation}
 \langle\hat n_\mathrm{j}(x,\omega)\rangle=\frac{\int_{-\infty}^\infty\rho_\mathrm{NL,j}(x,\omega,x')\langle\hat\eta(x',\omega)\rangle dx'}{\int_{-\infty}^\infty\rho_\mathrm{NL,j}(x,\omega,x')dx'}.
 \label{eq:photonnumbers}
\end{equation}
In contrast to purely dielectric media,
the nonlocal densities of states (NLDOSs)
$\rho_\mathrm{NL,j}(x,\omega,x')$ now include additional
terms originating from magnetic field-matter interactions.
The generalized NLDOSs are given by
%
\begin{equation}
 \Scale[0.94]{
 \begin{array}{l}
 \rho_\mathrm{NL,e}(x,\omega,x')\\[6pt]
 \displaystyle=\frac{2\omega^3}{\pi c^4S}\Big[\varepsilon_\mathrm{i}(x',\omega)|G_\mathrm{ee}(x,\omega,x')|^2\!+\!\mu_\mathrm{i}(x',\omega)|G_\mathrm{em}(x,\omega,x')|^2\Big],
 \end{array}
 }
 \label{eq:enldos}
\end{equation}
\begin{equation}
 \Scale[0.94]{
 \begin{array}{l}
 \rho_\mathrm{NL,m}(x,\omega,x')\\[6pt]
 \displaystyle=\frac{2\omega^3}{\pi c^4S}\Big[\varepsilon_\mathrm{i}(x',\omega)|G_\mathrm{me}(x,\omega,x')|^2\!+\!\mu_\mathrm{i}(x',\omega)|G_\mathrm{mm}(x,\omega,x')|^2\Big],
 \end{array}
 }
 \label{eq:hnldos}
\end{equation}
\begin{equation}
 \Scale[0.94]{
 \begin{array}{l}
 \rho_\mathrm{NL,tot}(x,\omega,x')\\[6pt]
 \displaystyle=\frac{|\varepsilon(x,\omega)|}{2}\rho_\mathrm{NL,e}(x,\omega,x')+\frac{|\mu(x,\omega)|}{2}\rho_\mathrm{NL,m}(x,\omega,x').
 \end{array}
 }
 \label{eq:unldos}
\end{equation}

The local densities of states (LDOSs) also existing in the denominator
of Eq.~\eqref{eq:photonnumbers} are given in terms of the NLDOSs as
\begin{equation}
 \rho_\mathrm{j}(x,\omega)=\int_{-\infty}^\infty\rho_\mathrm{NL,j}(x,\omega,x')dx'.
 \label{eq:LDOSs}
\end{equation}
The electric and magnetic LDOSs in Eq.~\eqref{eq:LDOSs}
with $j\in\{\mathrm{e,m}\}$ are related to the imaginary parts of the respective
Green's functions $G_\mathrm{ee}(x,\omega,x)$ and $G_\mathrm{mm}(x,\omega,x)$ as
\begin{equation}
 \rho_j(x,\omega)=\frac{2\omega}{\pi c^2S}\mathrm{Im}[G_{jj}(x,\omega,x)].
\end{equation}
Note that the electric and magnetic LDOSs are directly given by the imaginary parts of the
Green's functions $G_\mathrm{ee}(x,\omega,x)$ and $G_\mathrm{mm}(x,\omega,x)$
even though the NLDOSs in Eqs.~\eqref{eq:enldos} and \eqref{eq:hnldos}
also depend on the Green's functions $G_\mathrm{em}(x,\omega,x)$
and $G_\mathrm{me}(x,\omega,x)$. This manifests the
intimate coupling of the four Green's functions.
Note that the obtained LDOSs are equivalent to those
obtained by using the conventional fluctuational electrodynamics
in the case of normal incidence
\cite{Joulain2003,Joulain2005,Narayanaswamy2010}.

The position-dependent photon-ladder operators contributing to the
effective photon-number expectation values in Eq.~\eqref{eq:photonnumbers}
can be obtained by the same procedure as that presented
for dielectric media in Ref.~\citenum{Partanen2014a}.
The only exception is that we now have two commuting source field operators
$\hat f_\mathrm{e}(x,\omega)$ and $\hat f_\mathrm{m}(x,\omega)$
instead of a single operator $\hat f(x,\omega)$.
The resulting expression for the ladder operators reads
\begin{align}
 \hat a_\mathrm{j}(x,\omega) &=\frac{1}{\sqrt{\rho_\mathrm{j}(x,\omega)}}\int_{-\infty}^\infty
 \Big[\sqrt{\rho_\mathrm{NL,j,e}(x,\omega,x')}\hat f_\mathrm{e}(x',\omega)\nonumber\\
 &\hspace{0.5cm}+\sqrt{\rho_\mathrm{NL,j,m}(x,\omega,x')}\hat f_\mathrm{m}(x',\omega)\Big]dx',
 \label{eq:annihilation}
\end{align}
where $\rho_\mathrm{NL,j,e}(x,\omega,x')$ and $\rho_\mathrm{NL,j,m}(x,\omega,x')$
with $j\in\{\mathrm{e,m}\}$ denote, respectively, the first and the second
terms of Eqs.~\eqref{eq:enldos} and \eqref{eq:hnldos}.
The total NLDOS terms $\rho_\mathrm{NL,tot,e}(x,\omega,x')$ and $\rho_\mathrm{NL,tot,m}(x,\omega,x')$
are calculated by using Eq.~\eqref{eq:unldos} with the corresponding terms
in the electric and magnetic NLDOSs.

As in the case of dielectric media \cite{Partanen2014a,Partanen2015a}, we apply the general definition
of the quantum optical Poynting vector operator
as a normal ordered operator in
terms of the positive and negative frequency parts of the electric 
and magnetic field operators, given by
$\hat{S}(x,t)=:\!\hat{E}(x,t)\hat{H}(x,t)\!:=\hat{E}^-(x,t)\hat{H}^+(x,t)+\hat{H}^-(x,t)\hat{E}^+(x,t)$
\cite{Loudon2000}.
Substituting the electric and magnetic field operators in
Eqs.~\eqref{eq:efield} and \eqref{eq:hfield} into the Poynting vector
definition and taking the expectation value results in
\begin{equation}
 \langle\hat S(x,t)\rangle_\omega=\hbar\omega v(x,\omega)\int_{-\infty}^\infty\rho_\mathrm{IF}(x,\omega,x')\langle\hat\eta(x',\omega)\rangle dx'
 \label{eq:poynting1},
\end{equation}
where $v(x,\omega)=c/n_\mathrm{r}(x,\omega)$ is the energy propagation velocity,
$n_\mathrm{r}(x,\omega)$ is the real part of the refractive index, and the quantity
$\rho_\mathrm{IF}(x,\omega,x')$, introduced for nonmagnetic media in
Ref.~\citenum{Partanen2015a}, is referred to as the interference
density of states (IFDOS) and it is, in the present case, given by
\begin{align}
 &\rho_\mathrm{IF}(x,\omega,x')\nonumber\\
 & =\frac{2\omega^2 n_\mathrm{r}(x,\omega)}{\pi c^4S}\Big[\varepsilon_\mathrm{i}(x',\omega)\mathrm{Re}\Big(i\omega G_\mathrm{ee}(x,\omega,x') G_\mathrm{me}^*(x,\omega,x')\Big)\nonumber\\
 &\hspace{0.5cm}+\mu_\mathrm{i}(x',\omega)\mathrm{Re}\Big(i\omega G_\mathrm{mm}(x,\omega,x')G_\mathrm{em}^*(x,\omega,x')\Big)\Big]
 \label{eq:ifdos}.
\end{align}
The integral of the IFDOS with respect to $x'$ is always zero as required, e.g. by the fact
that in a medium in thermal equilibrium, there is no net energy flow \cite{Partanen2015a}.
In addition, it is important to note that
the total Poynting vector expectation value in Eq.~\eqref{eq:poynting1} is always continuous
at interfaces, which is necessary due to the conservation of energy.
Note that, in the QFED framework, it is also possible to apply the density
of states concepts to express the Poynting vector
expectation value in Eq.~\eqref{eq:poynting1}
in terms of the left and right propagating photon-number expectation values
by using the procedure described in Ref.~\citenum{Partanen2015a}.

\subsection{Macroscopic emission and absorption operators and thermal balance}

A particularly insightful view of the effective photon numbers is provided by their connection
to local thermal balance between the field and matter studied in
the case of nonmagnetic media in Ref.~\citealp{Partanen2014a}.
Here we present the corresponding thermal balance equation in the case
of general media including field-matter interactions through the magnetic field.
First, we present the emission and absorption operators $\hat Q_\mathrm{em}(x,t)$ and
$\hat Q_\mathrm{abs}(x,t)$ that capture the macroscopic nature of the material
layers that are assumed to act as constant memoryless reservoirs.
In terms of the field and the noise current operators, the normal
ordered emission and absorption operators
$\hat Q_\mathrm{em}(x,t)$ and $\hat Q_\mathrm{abs}(x,t)$ are given by
\begin{align}
 \hat Q_\mathrm{em}(x,t)
 &=-:\hat J_\mathrm{e}(x,t)\hat E(x,t):
 -:\hat J_\mathrm{m}(x,t)\hat H(x,t):,
 \label{eq:emissionop}\\
 \hat Q_\mathrm{abs}(x,t)
 &=:\hat J_\mathrm{e,abs}(x,t)\hat E(x,t):
 +:\hat J_\mathrm{m,abs}(x,t)\hat H(x,t):.
 \label{eq:absorptionop}
\end{align}
The second terms describe the field-matter interactions through
the magnetic field and they are not present in the formalism
for dielectrics. Whereas the emission current operators
$\hat J_\mathrm{e}(x,t)$ and $\hat J_\mathrm{m}(x,t)$
directly present the field sources as shown in frequency domain in
Eq.~\eqref{eq:HelmholtzE2} and \eqref{eq:HelmholtzB2},
the absorption current operators $\hat J_\mathrm{e,abs}(x,t)$ and
$\hat J_\mathrm{m,abs}(x,t)$ describe secondary
currents that are induced by the electric and magnetic fields.
Therefore, the operators $\hat J_\mathrm{e}(x,t)$ and $\hat J_\mathrm{m}(x,t)$
can be seen to correspond to the classical free current densities
whereas the operators $\hat J_\mathrm{e,abs}(x,t)$ and
$\hat J_\mathrm{m,abs}(x,t)$
correspond to the classical bound current densities \cite{Jackson1999}.
The operators $\hat J_\mathrm{e,abs}(x,t)$ and
$\hat J_\mathrm{m,abs}(x,t)$
are written in the spectral domain as
$\hat J_\mathrm{e,abs}^+(x,\omega)=-i\omega\varepsilon_0\chi_\mathrm{e}(x,\omega)\hat E^+(x,\omega)$
and 
$\hat J_\mathrm{m,abs}^+(x,\omega)=-i\omega\mu_0\chi_\mathrm{m}(x,\omega)\hat H^+(x,\omega)$,
where $\chi_\mathrm{e}(x,\omega)=\varepsilon(x,\omega)-1$ and $\chi_\mathrm{m}(x,\omega)=\mu(x,\omega)-1$
are the electric and magnetic susceptibilities of the medium.
The total current density operators are then given by
$\hat J_\mathrm{e,tot}(x,t)=\hat J_\mathrm{e}(x,t)+\hat J_\mathrm{e,abs}(x,t)$
and
$\hat J_\mathrm{m,tot}(x,t)=\hat J_\mathrm{m}(x,t)+\hat J_\mathrm{m,abs}(x,t)$.

The net emission operator $\hat Q(x,t)=\hat Q_\mathrm{em}(x,t)-\hat Q_\mathrm{abs}(x,t)$,
whose expectation value equals the divergence of the Poynting vector
expectation value in Eq.~\eqref{eq:poynting1} \cite{Partanen2014a},
is given in terms of the electric and magnetic field operators and the total current density operators as
$\hat Q(x,t)=:\hat J_\mathrm{e,tot}(x,t)\hat E(x,t):+:\hat J_\mathrm{m,tot}(x,t)\hat H(x,t):$.
In the case of dielectrics,
the spectral component of the expectation value of the net emission operator
can be written in terms of the electric LDOS and the effective electric field photon number
expectation value \cite{Partanen2014a}.
In the present case, we obtain the net emission expectation value
in terms of both the electric and magnetic LDOSs
and the effective electric and magnetic field photon numbers as
\begin{align}
 &\langle\hat Q(x,t)\rangle_\omega\nonumber\\
 & =\hbar\omega^2\varepsilon_\mathrm{i}(x,\omega)\rho_\mathrm{e}(x,\omega)[\langle\hat\eta(x,\omega)\rangle-\langle\hat n_\mathrm{e}(x,\omega)\rangle]\nonumber\\
&\hspace{0.5cm}+\hbar\omega^2\mu_\mathrm{i}(x,\omega)\rho_\mathrm{m}(x,\omega)[\langle\hat\eta(x,\omega)\rangle-\langle\hat n_\mathrm{m}(x,\omega)\rangle].
 \label{eq:divP}
\end{align}
At global thermal equilibrium, the effective electric and magnetic
photon-number expectation values $\langle\hat n_\mathrm{e}(x,\omega)\rangle$
and $\langle\hat n_\mathrm{m}(x,\omega)\rangle$ both
reach the source field value $\langle\hat\eta(x,\omega)\rangle=\langle\hat \eta_0\rangle$ when
the net emission rate in Eq.~\eqref{eq:divP} becomes zero.
In resonant systems where the energy exchange is dominated by a narrow frequency band,
condition $\langle\hat Q(x,t)\rangle_\omega=0$ can be used
to determine the approximate steady state temperature of a weakly interacting resonant
particle \cite{Bohren1998}.

\section{\label{sec:results}Examples}

\begin{figure}
\includegraphics[width=0.48\textwidth]{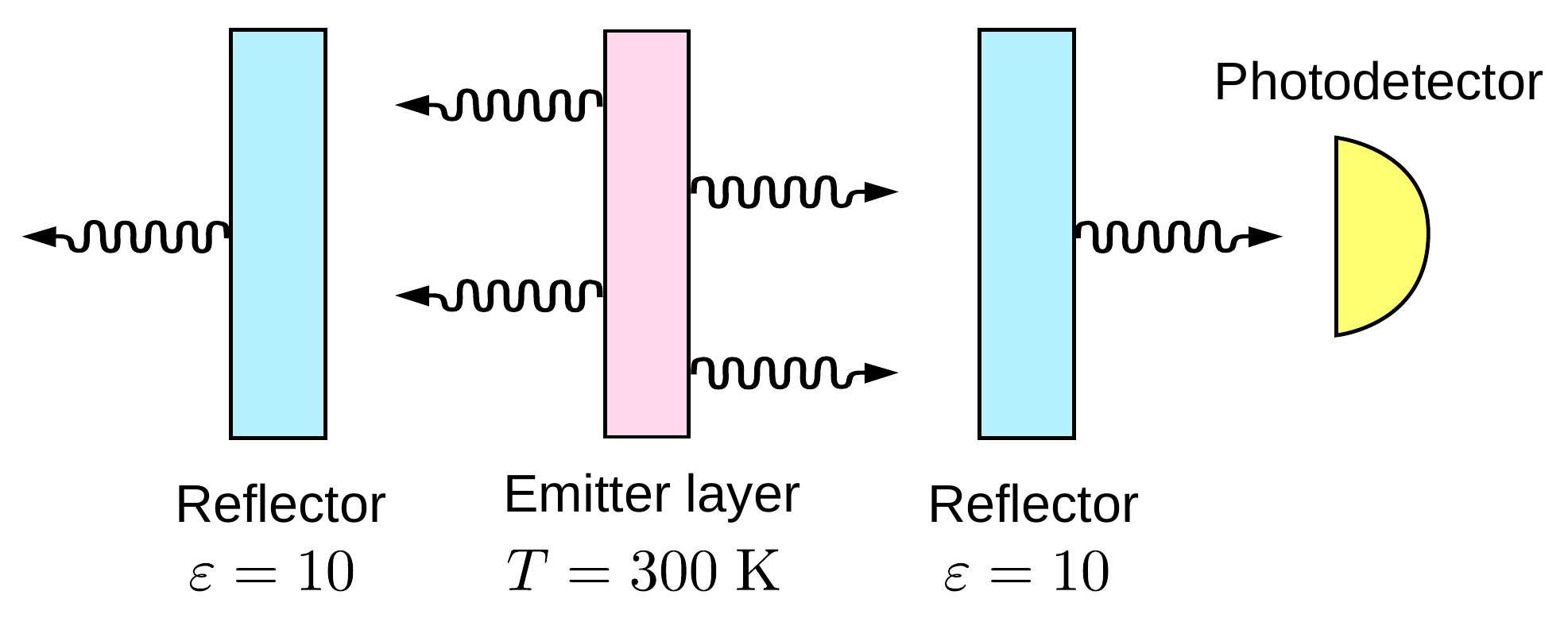}
\caption{\label{fig:emittersetup}(Color online) Schematic illustration of
the setup for measuring the emissivity of electric and magnetic
emitter layers placed in the middle of the vacuum cavity formed
by two reflectors. The
electric and magnetic LDOSs are position-dependent inside
the cavity which results in significantly different emissivities
of electric and magnetic emitters as detected by a photodetector
outside the cavity.}
\end{figure}

To investigate the physical implications and advantages
of the concepts presented in Sec.~\ref{sec:theory}
we apply the methods to study the emissivities
of electric and magnetic emitters placed in the middle of a vacuum cavity
to show that it is possible to directly probe
and demonstrate the essentially independent
nature of the electric and magnetic source terms.

In the example, a thin heated material layer which interacts with the
electromagnetic field through electric or magnetic interaction terms
is placed in the middle of
a 10 $\mu$m thick vacuum cavity as illustrated in Fig.~\ref{fig:emittersetup}.
The relative permittivity and permeability of the lossless 1 $\mu$m thick cavity walls are
$\varepsilon=10$ and $\mu=1$, resulting in the cavity wall power reflection coefficient
$R=0.64$ for the second cavity resonance with energy $\hbar\omega=0.119$ eV ($\lambda=10.4$ $\mu$m).
We focus on the second resonance since it exhibits a node for the electric field
and anti-node for the magnetic field in the middle of the cavity.

\begin{figure}
\includegraphics[width=0.45\textwidth]{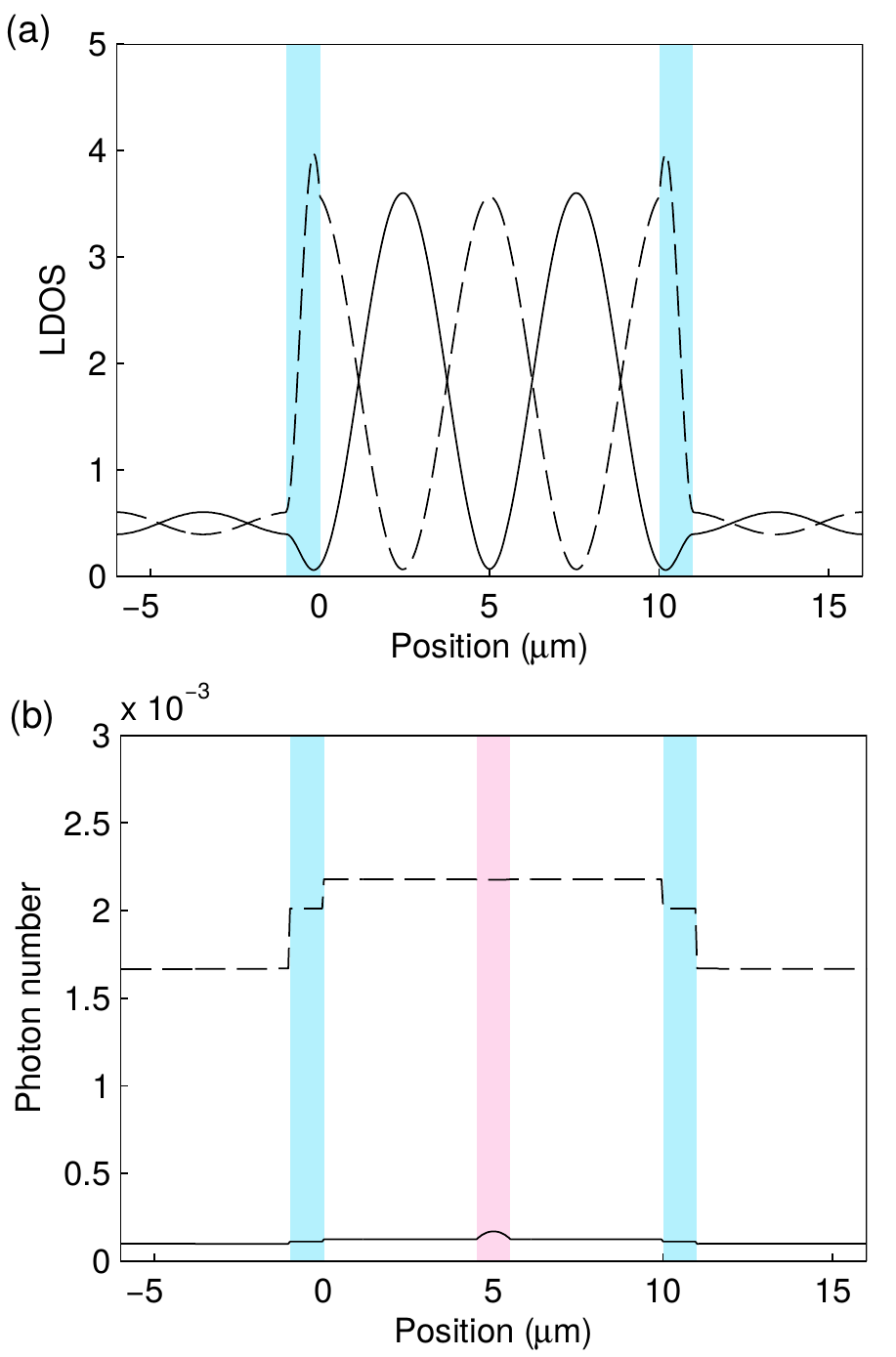}
\caption{\label{fig:emittergraphs}(Color online) (a) Electric (solid line)
and magnetic (dashed line) LDOSs in the vicinity of a vacuum cavity with
relative wall permittivity $\varepsilon=10$ and permeability $\mu=1$
for the second resonant energy $\hbar\omega=0.119$ eV ($\lambda=10.4$ $\mu$m).
(b) Effective photon number of the total
electromagnetic field for the above photon energy in the case of electric
(solid line) and magnetic (dashed line)
emitter layers at temperature $T=300$ K placed in the middle of the cavity.
The electric emitter layer has relative permittivity $\varepsilon=1.1+0.1i$
and permeability $\mu=1$ and the magnetic emitter layer has relative
permittivity $\varepsilon=1$ and permeability $\mu=1.1+0.1i$.
The width of the cavity is 10 $\mu$m and the thickness of the
cavity walls and the emitter layers is 1 $\mu$m.
The LDOSs are given in the units of $2/(\pi c S)$.}
\end{figure}

Figure \ref{fig:emittergraphs}(a)
shows the electric and magnetic LDOSs for the second resonant photon energy.
Inside the cavity, the electric LDOS has two maxima and the magnetic LDOS,
respectively, has two minima.
In the middle of the cavity, the electric LDOS obtains its minimum value
which is close to zero. The magnetic LDOS instead obtains its maximum
value at the same point. The LDOSs determine the local field-matter
interactions as seen in the net emission rate in Eq.~\eqref{eq:divP}.

Figure \ref{fig:emittergraphs}(b)
shows the effective photon number of
the total electromagnetic field for the second resonant
energy when a 1 $\mu$m thick electric or magnetic
emitter layer with temperature $T=300$ K is placed in the middle
of the cavity. The electric emitter
layer has a relative permittivity $\varepsilon=1.1+0.1i$ and a permeability
$\mu=1$ and the magnetic emitter layer, respectively, has a relative
permittivity $\varepsilon=1$ and a permeability $\mu=1.1+0.1i$.
The photon number is piecewise constant in all lossless media
in the geometry. It can be seen that
the magnitude of the photon number is significantly
larger in the case of the magnetic emitter layer
due to the different emissivities following from the
electric field node and the magnetic field
anti-node at the position of the emitter layer.
This behavior, which manifests the magnetic Purcell effect,
should be directly experimentally measurable by
detecting the normally emitted field outside the cavity.
The experimental verification of the phenomenon would
therefore demonstrate the essentially independent nature of the
polarization and magnetization source terms.

To experimentally verify the existence of the magnetic Purcell effect
and the need for the second noise source term, it would be necessary to construct
a cavity system with a magnetically interacting emitter layer, in addition to
a conventional electrically interacting one. The emitter layer might need to be
constructed using engineered materials \cite{Shalaev2007,Tighineanu2014}.
The cavity itself could consist of almost any kind of transparent materials
or vacuum, and partly reflecting walls allowing an appropriate amount of emission
to escape from the cavity. The detection of the light could then be carried out
by using any frequency and direction selective setup, such as the setup
presented by Marquier \emph{et al.} \cite{Marquier2004}.
In agreement with the Kirchoff's law of radiation, a complementary setup could
also be constructed in reverse, measuring the power absorbed by the Purcell enhanced absorber.

\section{\label{sec:conclusions}Conclusions}
\vspace{-0.1cm}

In conclusion, we have formulated a generalized QFED
noise operator formalism that is able to consistently
describe the effective photon number,
field-matter interactions, and the formation of thermal
balance in nonequilibrium conditions
in general isotropic media. It was shown that
two commuting bosonic noise operators are
needed to describe the field sources
for a single polarization
in order to maintain the well-known canonical
commutation relations for the field operators.
The two bosonic noise operators
are directly related to the electric
and magnetic field-matter interactions in the medium.

We have also used the model to predict how thermal emission from
electric or magnetic emitters changes in a configuration where
the emitters are located in an optical cavity.
The results suggest that it is possible to design
a conceptually straightforward experimental setup
to differentiate between the two independent noise
sources with fundamentally different origin.

\vspace{-0.1cm}
\begin{acknowledgments}
This work has in part been funded by the Academy of Finland and the Aalto Energy Efficiency Research Programme.
\end{acknowledgments}

\vspace{-0.5cm}
\appendix

\section{\label{apx:green}Green's functions}

\subsection{Multi-interface reflection and transmission coefficients}

We first define the conventional single interface electric and magnetic
field reflection and transmission
coefficients $r_\mathrm{e}$, $r_\mathrm{m}$, $t_\mathrm{e}$,
$t_\mathrm{m}$. The materials on the left and right have relative permittivities and permeabilities
$\varepsilon_1$, $\mu_1$, $\varepsilon_2$, and $\mu_2$, and refractive indices
$n_1=\sqrt{\varepsilon_1\mu_1}$ and $n_2=\sqrt{\varepsilon_2\mu_2}$.
For left normal incidence the reflection and transmission coefficients are given by
\begin{align}
 r_\mathrm{e} &=\frac{\mu_2n_1-\mu_1n_2}{\mu_2n_1+\mu_1n_2},\hspace{0.5cm}
 t_\mathrm{e} =\frac{2\mu_2n_1}{\mu_2n_1+\mu_1n_2},\nonumber\\
 r_\mathrm{m} &=\frac{\varepsilon_2n_1-\varepsilon_1n_2}{\varepsilon_2n_1+\varepsilon_1n_2},\hspace{0.5cm}
 t_\mathrm{m} =\frac{2\varepsilon_2n_1}{\varepsilon_2n_1+\varepsilon_1n_2},
 \label{eq:fresnel}
\end{align}
The reflection and transmission coefficients for right incidence $r_\mathrm{e}'$,
$r_\mathrm{m}'$, $t_\mathrm{e}'$, and $t_\mathrm{m}'$ are
obtained by switching the indices 1 and 2 in the expressions of
$r_\mathrm{e}$, $r_\mathrm{m}$, $t_\mathrm{e}$, and $t_\mathrm{m}$.

The multi-interface geometry is defined by interface positions $x_l$, $l=1,2,...,N$
separating material layers with relative permittivities and permeabilities $\varepsilon_l$ and
$\mu_l$, refractive indices $n_l$ and wavenumbers $k_l$, where
$l=1,2,...,N+1$. The layer thicknesses are denoted by $d_l=x_l-x_{l-1}$, where $l=2,...,N$.
The multi-interface reflection and transmission coefficients $\mathcal{R}_{l,j}$
and $\mathcal{T}_{l,j}$,
which account for the multiple reflections in different medium layers,
are recursively given in terms of the single interface reflection
and transmission coefficients as
\begin{align}
 \mathcal{R}_{l,j} & =\frac{r_{l,j}+\mathcal{R}_{l+1,j}e^{2ik_{l+1}d_{l+1}}}{1+r_{l,j}\mathcal{R}_{l+1,j}e^{2ik_{l+1}d_{l+1}}}\label{eq:Rl}\\
 \mathcal{T}_{l,j} & =\frac{t_{l,j}\nu_{l+1,j}}{\nu_{l,j}(1-\mathcal{R}_{l-1,j}'r_{l,j}e^{2ik_ld_l})}\label{eq:Tl},
\end{align}
where $l=1,2,...,N$, $j\in\{\mathrm{e},\mathrm{m}\}$, $\nu_{l,j}=1/(1-\mathcal{R}_{l-1,j}'\mathcal{R}_{l,j} e^{2ik_ld_l})$, and
$\mathcal{R}_{0,j}'=\mathcal{R}_{N+1,j}=0$. As in the case of single interface coefficients
in Eq.~\eqref{eq:fresnel} the primed coefficients denote the coefficients for right incidence.
The layers are indexed such that $\mathcal{R}_{l,j}'$
corresponds to the same interface as $\mathcal{R}_{l,j}$. The propagation coefficient
for a material layer $l$ of thickness $d_l$ is given by $e^{ik_ld_l}$,
the transmission coefficient $\mathcal{T}_{l,l',j}$
from layer $l'$ to layer $l>l'+1$ is recursively given by
$\mathcal{T}_{l,l',j}=\mathcal{T}_{l-1,l',j}\mathcal{T}_{l-1,j}e^{ik_{l-1}d_{l-1}}$
with $\mathcal{T}_{l'+1,l',j}=\mathcal{T}_{l',j}$, and the transmission coefficient
$\mathcal{T}_{l,l',j}'$ from layer $l'$ to layer $l<l'-1$ is given by
$\mathcal{T}_{l,l',j}'=\mathcal{T}_{l+1,l',j}'\mathcal{T}_{l,j}'e^{ik_{l+1}d_{l+1}}$
with $\mathcal{T}_{l'-1,l',j}'=\mathcal{T}_{l'-1,j}'$.

\subsection{Green's functions for layered structures}

We write the electric and magnetic Green's functions $G_\mathrm{ee}(x,\omega,x')$
and $G_\mathrm{mm}(x,\omega,x')$ for a general layered structure
in terms of the scaled Green's functions $\xi_j(x,\omega,x')$ defined below as
\begin{align}
 G_\mathrm{ee}(x,\omega,x') &=\mu(x',\omega)\xi_\mathrm{e}(x,\omega,x')\label{eq:greendefe}\\
 G_\mathrm{mm}(x,\omega,x') &=\varepsilon(x',\omega)\xi_\mathrm{m}(x,\omega,x').\label{eq:greendefm}
\end{align}

In the following, the source point $x'$ is located in layer $l'$ ($x_{l'-1}<x'<x_{l'}$)
and field point $x$ is located in layer $l$ ($x_{l-1}<x<x_{l}$) with $x_0=-\infty$
and $x_{N+1}=\infty$.
In the source layer ($l=l'$), the scaled Green's function has three components as
\begin{align}
 &\xi_{l=l',j}(x,\omega,x')\nonumber\\
 &=\xi_{0,l'}(x,\omega,x')+\xi_{+,l',j}(x,\omega,x')+\xi_{-,l',j}(x,\omega,x').
\end{align}
The component $\xi_{0,l'}(x,\omega,x')$ is the homogeneous space solution and the
components $\xi_{+,l',j}(x,\omega,x')$ and $\xi_{-,l',j}(x,\omega,x')$ describe the right and left
propagating fields due to the reflections at the interfaces.
The homogeneous space solution is given by
\begin{equation}
 \xi_{0,l'}(x,\omega,x')=\frac{ie^{ik_{l'}|x-x'|}}{2k_{l'}}
\end{equation}
and the right propagating reflection originating component is written as
\begin{equation}
 \Scale[0.96]{
 \begin{array}{l}
 \xi_{+,l',j}(x,\omega,x')\\[6pt]
 \displaystyle= e^{ik_{l'}(x-x_{l'-1})}\xi_{0,l'}(x_{l'-1},\omega,x')\mathcal{R}_{l'-1,j}'\\[6pt]
 \displaystyle\hspace{0.5cm}\times\sum_{m=0}^\infty(\mathcal{R}_{l'-1,j}'\mathcal{R}_{l',j}e^{2ik_{l'}d_{l'}})^m\\[6pt]
 \displaystyle\hspace{0.5cm}+e^{ik_{l'}(x-x_{l'-1})}\xi_{0,l'}(x_{l'},\omega,x')\mathcal{R}_{l'-1,j}'\mathcal{R}_{l',j}e^{ik_jd_j}\\[6pt]
 \displaystyle\hspace{0.5cm}\times\sum_{m=0}^\infty(\mathcal{R}_{l'-1,j}'\mathcal{R}_{l',j}e^{2ik_{l'}d_{l'}})^m\\[6pt]
 \displaystyle=e^{ik_{l'}(x-x_{l-1})}\frac{ie^{ik_{l'}(x'-x_{l'-1})}}{2k_{l}}\nu_{l',j}\mathcal{R}_{l'-1,j}'\\[6pt]
 \displaystyle\hspace{0.5cm}+e^{ik_{l'}(x-x_{l'-1})}\frac{ie^{ik_{l'}(x_{l'}-x')}}{2k_{l'}}e^{ik_{l'}d_{l}}\nu_{l',j}\mathcal{R}_{l'-1}'\mathcal{R}_{l'}\\[6pt]
 \displaystyle=\frac{i}{2k_{l'}}\nu_{l',j}\mathcal{R}_{l'-1,j}'(e^{ik_{l'}(x+x'-2x_{l'-1})}+\mathcal{R}_{l',j}e^{ik_{l'}(x-x'+2d_{l'})}).
 \end{array}
 }
\end{equation}
The first term describes the field component incident from the source point to the left and the second
term describes the field component incident from the source point to the right.
The factor $\nu_{l',j}=1/(1-\mathcal{R}_{l'-1,j}'\mathcal{R}_{l',j} e^{2ik_{l'}d_{l'}})$
arises from the series accounting for the
multiple reflections inside the source layer.
Respectively, the left propagating  reflection originating component is written as
\begin{equation}
 \Scale[0.95]{
 \begin{array}{l}
 \xi_{-,l',j}(x,\omega,x')\\[6pt]
 \displaystyle= e^{-ik_{l'}(x-x_{l'})}\xi_{0,l'}(x_{l'},\omega,x')\mathcal{R}_{l',j}\\[6pt]
 \displaystyle\hspace{0.5cm}\sum_{m=0}^\infty(\mathcal{R}_{l'-1,j}'\mathcal{R}_{l',j}e^{2ik_{l'}d_{l'}})^m\\[6pt]
 \displaystyle\hspace{0.5cm}+e^{-ik_{l'}(x-x_{l'})}\xi_{0,l'}(x_{l'-1},\omega,x')\mathcal{R}_{l'-1,j}'\mathcal{R}_{l',j}e^{ik_{l'}d_{l'}}\\[6pt]
 \displaystyle\hspace{0.5cm}\sum_{m=0}^\infty(\mathcal{R}_{l'-1,j}'\mathcal{R}_{l',j}e^{2ik_{l'}d_{l'}})^m\\[6pt]
 \displaystyle=e^{-ik_{l'}(x-x_{l'})}\frac{ie^{ik_{l'}(x_{l'}-x')}}{2k_{l'}}\nu_{l',j}\mathcal{R}_{l',j}\\[6pt]
 \displaystyle\hspace{0.5cm}+e^{-ik_{l'}(x-x_{l'})}\frac{ie^{ik_{l'}(x'-x_{l'-1})}}{2k_{l'}}e^{ik_{l'}d_{l'}}\nu_{l',j}\mathcal{R}_{l'-1,j}'\mathcal{R}_{l',j}\\[6pt]
 \displaystyle=\frac{i}{2k_{l'}}\nu_{l',j}\mathcal{R}_{l',j}(e^{-ik_{l'}(x+x'-2x_{l'})}+\mathcal{R}_{l'-1,j}'e^{-ik_{l'}(x-x'-2d_{l'})}).
 \end{array}
 }
\end{equation}
Therefore, the total scaled Green's function is given in the source layer by
\begin{align}
 &\xi_{l=l',j}(x,\omega,x')\nonumber\\
 &=\frac{i}{2k_{l'}}\Big(e^{ik_{l'}|x-x'|}+\nu_{l',j}\mathcal{R}_{l',j}
 [e^{-ik_{l'}(x+x'-2x_{l'})}\nonumber\\
 &\hspace{0.5cm}+\mathcal{R}_{l'-1,j}'e^{-ik_{l'}(x-x'-2d_{l'})}]+\nu_{l',j}\mathcal{R}_{l'-1,j}'\nonumber\\
 &\hspace{0.5cm}\times[e^{ik_{l'}(x+x'-2x_{l'-1})}+\mathcal{R}_{l',j}e^{ik_{l'}(x-x'+2d_{l'})}]\Big).
\end{align}
Writing the scaled Green's functions in other layers is even more straightforward
and, as a result, the scaled Green's functions are given in the cases $l>l'$
and $l<l'$ by
\begin{align}
 &\xi_{l>l',j}(x,\omega,x')\nonumber\\
 &=\frac{i}{2k_{l'}}\mathcal{T}_{l,l',j}\Big(e^{ik_{l'}(x_{l'}-x')}+\nu_{l',j}\mathcal{R}_{l'-1,j}'
 [e^{ik_{l'}(x'-x_{l'-1}+d_{l'})}\nonumber\\
 &\hspace{0.5cm}+\mathcal{R}_{l',j}e^{ik_{l'}(2d_{l'}-x'+x_{l'})}]\Big)\nonumber\\
 &\times\Big(e^{ik_{l}(x-x_{l-1})}+\mathcal{R}_{l,j}e^{-ik_{l}(x-x_{l-1}-2d_{l})}\Big),
\end{align}
\begin{align}
 &\xi_{l<l',j}(x,\omega,x')\nonumber\\
 &=\frac{i}{2k_{l'}}\mathcal{T}_{l,l',j}'\Big(e^{ik_{l'}(x'-x_{l'-1})}\!+\!\nu_{l',j}\mathcal{R}_{l',j}
 [e^{-ik_{l'}(x'-x_{l'-1}-2d_{l'})}\nonumber\\
 &\hspace{0.5cm}+\mathcal{R}_{l'-1,j}'e^{ik_{l'}(x'-x_{l'-1}+2d_{l'})}]\Big)\nonumber\\
 &\hspace{0.5cm}\times\Big(e^{-ik_{l}(x-x_{l})}+\mathcal{R}_{l-1,j}'e^{ik_{l}(x-x_{l-1}+d_{l})}\Big).
\end{align}

\section{\label{apx:commutation}Canonical commutation relations of fields}
Here we determine the normalization factors $j_\mathrm{0,e}(x,\omega)$
and $j_\mathrm{0,m}(x,\omega)$ of the noise current operators by requiring
that the vector potential and electric field operators obey the canonical
commutation relation $[\hat{A}(x,t),\hat{E}(x',t)]=-i\hbar/(\varepsilon_0S)\delta(x-x')$
\cite{Barnett1995,Barnett1996,Matloob1995}.
The vector potential and electric field operators are given in the frequency domain by
\begin{align}
 &\hat{A}^+(x,\omega)\nonumber\\
 &= \mu_0\int_{-\infty}^\infty j_\mathrm{0,e}(x',\omega) G_\mathrm{ee}(x,\omega,x')\hat f_\mathrm{e}(x',\omega)dx'\nonumber\\
 &\hspace{0.5cm}+\frac{1}{ic}\int_{-\infty}^\infty j_\mathrm{0,m}(x',\omega)G_\mathrm{em}(x,\omega,x')\hat f_\mathrm{m}(x',\omega)dx'.\label{eq:apxafield}\\
 &\hat{E}^+(x,\omega)\nonumber\\
 &= i\omega\mu_0\int_{-\infty}^\infty j_\mathrm{0,e}(x',\omega) G_\mathrm{ee}(x,\omega,x')\hat f_\mathrm{e}(x',\omega)dx'\nonumber\\
 &\hspace{0.5cm}+k_0\int_{-\infty}^\infty j_\mathrm{0,m}(x',\omega)G_\mathrm{em}(x,\omega,x')\hat f_\mathrm{m}(x',\omega)dx'.\label{eq:apxefield}
\end{align}
The frequency domain commutator is obtained by using the
frequency domain vector potential and electric field operators as
\begin{equation}
 \Scale[0.94]{
 \begin{array}{l}
 [\hat{A}^+\,^\dag(x,\omega),\hat{E}^+(x',\omega')]\\[6pt]
 \displaystyle=i\omega\mu_0^2\int_{-\infty}^\infty\int_{-\infty}^\infty j_\mathrm{0,e}(y,\omega)j_\mathrm{0,e}^*(y',\omega')G_\mathrm{ee}(x,\omega,y)\\[12pt]
 \displaystyle\hspace{0.5cm}\times G_\mathrm{ee}^*(x',\omega',y')[\hat{f}_\mathrm{e}^\dag(y,\omega),\hat{f}_\mathrm{e}(y',\omega')]dydy'\\[6pt]
 \displaystyle\hspace{0.5cm}-\frac{k_0}{ic}\int_{-\infty}^\infty\int_{-\infty}^\infty j_\mathrm{0,m}(y,\omega)j_\mathrm{0,m}^*(y',\omega')G_\mathrm{em}(x,\omega,y)\\[12pt]
 \displaystyle\hspace{0.5cm}\times G_\mathrm{em}^*(x',\omega',y')[\hat{f}_\mathrm{m}^\dag(y,\omega),\hat{f}_\mathrm{m}(y',\omega')]dydy'\\[6pt]
 \displaystyle=\frac{\delta(\omega-\omega')}{i\omega}\Bigg[\omega^2\mu_0^2\int_{-\infty}^\infty\!\!\!|j_\mathrm{0,e}(y,\omega)|^2G_\mathrm{ee}(x,\omega,y)G_\mathrm{ee}^*(x',\omega,y)dy\\[6pt]
 \displaystyle\hspace{0.5cm}+k_0^2\int_{-\infty}^\infty|j_\mathrm{0,m}(y,\omega)|^2G_\mathrm{em}(x,\omega,y)G_\mathrm{em}^*(x',\omega,y)dy\Bigg]\\[6pt]
 \displaystyle=-i\frac{4\pi\hbar\omega}{\varepsilon_0c^2S}\delta(\omega-\omega')\Bigg[k_0^2\int_{-\infty}^\infty[\varepsilon_\mathrm{i}(y,\omega)+C_1(y,\omega)]\\[12pt]
 \displaystyle\hspace{0.5cm}\times G_\mathrm{ee}(x,\omega,y)G_\mathrm{ee}^*(x',\omega,y)dy\\[0pt]
 \displaystyle\hspace{0.5cm}+k_0^2\int_{-\infty}^\infty\!\!\![\mu_\mathrm{i}(y,\omega)+C_2(y,\omega)]G_\mathrm{em}(x,\omega,y)G_\mathrm{em}^*(x',\omega,y)dy\Bigg]\\[6pt]
 \displaystyle=-i\frac{4\pi\hbar\omega}{\varepsilon_0c^2S}\delta(\omega-\omega')\mathrm{Im}[G_\mathrm{ee}(x,\omega,x')]\\[12pt]
 \displaystyle\hspace{0.5cm}-i\frac{4\pi\hbar\omega^3}{\varepsilon_0c^4S}\delta(\omega-\omega')\int_{-\infty}^\infty C(x,x',y,\omega)dy,
 \vspace{-0.3cm}
 \end{array}
 }
 \label{eq:commutatorderivation1}
\end{equation}
where we have first substituted the field operators in
Eqs.~\eqref{eq:apxafield} and \eqref{eq:apxefield}, then applied the commutation relations
of operators $\hat{f}_\mathrm{e}(y,\omega)$ and $\hat{f}_\mathrm{m}(y,\omega)$ after which
we have substituted $|j_\mathrm{0,e}(y,\omega)|^2=4\pi\hbar\omega^2\varepsilon_0[\varepsilon_\mathrm{i}(y,\omega)+C_1(y,\omega)]/S$ and
$|j_\mathrm{0,m}(y,\omega)|^2=4\pi\hbar\omega^2\mu_0[\mu_\mathrm{i}(y,\omega)+C_2(y,\omega)]/S$,
where $C_1(y,\omega)$ and $C_2(y,\omega)$ are
functions that will be determined at the end of the calculation.
These substitutions essentially just transform the undetermined factors
$|j_\mathrm{0,e}(y,\omega)|^2$ and $|j_\mathrm{0,m}(y,\omega)|^2$ to undetermined
factors $C_1(y,\omega)$ and $C_2(y,\omega)$.
The forms of the substitutions are chosen so that the functions $C_1(y,\omega)$ and
$C_2(y,\omega)$ can be shown to be zero at the end of the calculation.
In the final step in Eq.~\eqref{eq:commutatorderivation1},
we have applied the Green's function integral identity
in Eq.~\eqref{eq:greeneidentity} and denoted
\vspace{-0.1cm}
\begin{align}
 C(x,x',y,\omega) &=C_1(y,\omega)G_\mathrm{ee}(x,\omega,y)G_\mathrm{ee}^*(x',\omega,y)\nonumber\\
 &\hspace{0.5cm}+C_2(y,\omega)G_\mathrm{em}(x,\omega,y)G_\mathrm{em}^*(x',\omega,y).
\label{eq:factorfunction}
\end{align}
\vspace{-0.4cm}

Next, we present the time domain equal-time commutator.
The time domain operators are given in terms of the frequency domain operators as
\vspace{-0.1cm}
\begin{align}
 \hat{A}(x,t)=\frac{1}{2\pi}\int_0^\infty\hat{A}^+(x,\omega)e^{-i\omega t}d\omega+H.c.,\\
 \hat{E}(x,t)=\frac{1}{2\pi}\int_0^\infty\hat{E}^+(x,\omega)e^{-i\omega t}d\omega+H.c.,
\end{align}
where $H.c.$ denotes the Hermitian conjugate of the first term.
In the calculation of the commutator, we also use the general relation
$[\hat{A}^+(x,\omega),\hat{E}^+\,^\dag(x',\omega')]=[\hat{A}^+\,^\dag(x,\omega),\hat{E}^+(x',\omega')]$,
which relates the commutators of the conjugate terms.
By using Eq.~\eqref{eq:commutatorderivation1},
the time domain equal-time commutator is then given by
\begin{align}
 &[\hat{A}(x,t),\hat{E}(x',t)]\nonumber\\
 &=\frac{1}{4\pi^2}\int_0^\infty\int_0^\infty[\hat{A}^+\,^\dag(x,\omega),\hat{E}^+(x',\omega')]e^{-i(\omega-\omega')t}d\omega d\omega'\nonumber\\
 &\hspace{0.5cm}+\frac{1}{4\pi^2}\int_0^\infty\int_0^\infty[\hat{A}^+(x,\omega),\hat{E}^+\,^\dag(x',\omega')]e^{i(\omega-\omega')t}d\omega d\omega'\nonumber\\
 &=-i\frac{2\hbar}{\pi\varepsilon_0c^2S}\int_0^\infty\omega\mathrm{Im}[G_\mathrm{ee}(x,\omega,x')]d\omega\nonumber\\
 &\hspace{0.5cm}-i\frac{2\hbar}{\pi\varepsilon_0c^4S}\int_0^\infty\int_{-\infty}^\infty\omega^3C(x,x',y,\omega)dyd\omega\nonumber\\
 &=-\frac{\hbar}{\pi\varepsilon_0c^2S}\int_0^\infty\omega[G_\mathrm{ee}(x,\omega,x')-G_\mathrm{ee}^*(x,\omega,x')]d\omega\nonumber\\
 &\hspace{0.5cm}-i\frac{2\hbar}{\pi\varepsilon_0c^4S}\int_0^\infty\int_{-\infty}^\infty\omega^3C(x,x',y,\omega)dyd\omega\nonumber\\
 &=-\frac{\hbar}{\pi\varepsilon_0c^2S}\int_{-\infty}^\infty\omega G_\mathrm{ee}(x,\omega,x')d\omega\nonumber\\
 &\hspace{0.5cm}-i\frac{2\hbar}{\pi\varepsilon_0c^4S}\int_0^\infty\int_{-\infty}^\infty\omega^3C(x,x',y,\omega)dyd\omega.
 \label{eq:commutatorderivation2}
\end{align}
where we have used the relation $2i\mathrm{Im}(z)=z-z^*$, $z\in\mathbb{C}$,
and then, for the resulting conjugate term, the Green's function conjugation relation
$G_\mathrm{ee}^*(x,\omega,x')=G_\mathrm{ee}(x,-\omega,x')$ and the
change of variables $\omega\longrightarrow-\omega$. This allows
us to express the first integral term as an integral over the whole
real axis.

By substituting different Green's function terms into the first term in
the result of Eq.~\eqref{eq:commutatorderivation2},
it can be shown that only the homogeneous space solution term of the
Green's function contributes to the result.
Substituting the homogeneous space solution term of the Green's function into
the first term in Eq.~\eqref{eq:commutatorderivation2}, we obtain
\begin{align}
 &[\hat{A}(x,t),\hat{E}(x',t)]\nonumber\\
 &=-\frac{i\hbar}{\pi\varepsilon_0c^2S}\int_{-\infty}^\infty\omega\mu(x',\omega)\frac{e^{i\omega n(x',\omega)|x-x'|/c}}{2\omega n(x',\omega)/c}d\omega\nonumber\\
 &\hspace{0.5cm}-i\frac{2\hbar}{\pi\varepsilon_0c^4S}\int_0^\infty\int_{-\infty}^\infty\omega^3C(x,x',y,\omega)dyd\omega\nonumber\\
 &=-\frac{\hbar}{2\pi^2\varepsilon_0c^2S}\!\int_{-\infty}^\infty\!\!\!\!\omega\mu(x',\omega)\!\!\int_{-\infty}^\infty\!\!\frac{e^{ik(x-x')}}{k^2-\omega^2 n(x',\omega)^2/c^2}dkd\omega\nonumber\\
 &=-\frac{\hbar}{2\pi^2\varepsilon_0S}\int_{-\infty}^\infty e^{ik(x-x')}\int_{-\infty}^\infty\frac{\omega\mu(x',\omega)}{k^2c^2-\omega^2 n(x',\omega)^2}d\omega dk\nonumber\\
 &\hspace{0.5cm}-i\frac{2\hbar}{\pi\varepsilon_0c^4S}\int_0^\infty\int_{-\infty}^\infty\omega^3C(x,x',y,\omega)dyd\omega\nonumber\\
 &=-\frac{i\hbar}{\varepsilon_0S}\frac{1}{2\pi}\int_{-\infty}^\infty e^{ik(x-x')}dk\nonumber\\
 &\hspace{0.5cm}-i\frac{2\hbar}{\pi\varepsilon_0c^4S}\int_0^\infty\int_{-\infty}^\infty\omega^3C(x,x',y,\omega)dyd\omega\nonumber\\
 &=-\frac{i\hbar}{\varepsilon_0S}\delta(x\!-\!x')\!-\!i\frac{2\hbar}{\pi\varepsilon_0c^4S}\!\int_0^\infty\!\!\int_{-\infty}^\infty\!\!\!\omega^3C(x,x',y,\omega)dyd\omega,
 \label{eq:commutatorderivation3}
\end{align}
where we have applied the mathematical integral identities
in Eqs.~\eqref{eq:math1} and \eqref{eq:math2}
and the definition of the Dirac delta function.
From the final result, it follows that the second term
must be zero as the canonical commutation relation is known
to be given by $[\hat{A}(x,t),\hat{E}(x',t)]=-i\hbar/(\varepsilon_0S)\delta(x-x')$
\cite{Barnett1995,Barnett1996,Matloob1995}.
Therefore, as the integral of $C(x,x',y,\omega)$ must give zero
for all values of $x$ and $x'$ and as the position-dependence of
$C(x,x',y,\omega)$ comes directly from two linearly
independent Green's functions as presented in Eq.~\eqref{eq:factorfunction},
we must have $C_1(y,\omega)=C_2(y,\omega)=0$.
This condition then fixes the values of the normalization factors
$j_\mathrm{0,e}(x,\omega)$ and $j_\mathrm{0,m}(x,\omega)$ to
\begin{align}
 j_\mathrm{0,e}(x,\omega) &=\sqrt{4\pi\hbar\omega^2\varepsilon_0\varepsilon_\mathrm{i}(x,\omega)/S},\\
 j_\mathrm{0,m}(x,\omega) &=\sqrt{4\pi\hbar\omega^2\mu_0\mu_\mathrm{i}(x,\omega)/S},
\end{align}
which are unique apart from the possible phase factors.

\section{Mathematical identities}

\subsection{Green's function integral identities}

Here we derive the integral identities for the Green's function used in the final
step in the evaluation of Eq.~\eqref{eq:commutatorderivation1}.
The electric Green's function obeys the differential equation in Eq.~\eqref{eq:Greenee}
which we write by renaming the variables as
\begin{equation}
 \frac{\partial}{\partial y}\Big(\frac{\partial G_\mathrm{ee}(y,\omega,x)}{\mu(y,\omega)\partial y}\Big)+k_0^2\varepsilon(y,\omega)G_\mathrm{ee}(y,\omega,x)=-\delta(y-x).
\end{equation}
We multiply this with the conjugated Green's function $G_\mathrm{ee}^*(y,\omega,x')$ and integrate over $y$ to obtain
\begin{align}
 &\int_{-\infty}^\infty G_\mathrm{ee}^*(y,\omega,x')\frac{\partial}{\partial y}\Big(\frac{\partial G_\mathrm{ee}(y,\omega,x)}{\mu(y,\omega)\partial y}\Big)dy\nonumber\\
 &+k_0^2\int_{-\infty}^\infty\varepsilon(y,\omega)G_\mathrm{ee}^*(y,\omega,x')G_\mathrm{ee}(y,\omega,x)dy\nonumber\\
 &=-G_\mathrm{ee}^*(x,\omega,x').
\end{align}
Here, the first term can be integrated by parts accounting for the fact that the boundary term
becomes zero as the Green's function is exponentially decaying in lossy media and lossless
media can be studied in the limit of small losses. Therefore, we get
\begin{align}
 &-\int_{-\infty}^\infty\mu^*(y,\omega)\frac{\partial G_\mathrm{ee}(y,\omega,x)}{\mu(y,\omega)\partial y}\frac{\partial G_\mathrm{ee}^*(y,\omega,x')}{\mu^*(y,\omega)\partial y}dy\nonumber\\
 &+k_0^2\int_{-\infty}^\infty\varepsilon(y,\omega)G_\mathrm{ee}(y,\omega,x)G_\mathrm{ee}^*(y,\omega,x')dy\nonumber\\
 &=-G_\mathrm{ee}^*(x,\omega,x').
\end{align}
The first term can be expressed in terms of the Green's function $G_\mathrm{me}(x,\omega,x')$
by using Eq.~\eqref{eq:greenme}. Thus, we obtain
\begin{align}
 &-k_0^2\int_{-\infty}^\infty\mu^*(y,\omega)G_\mathrm{me}(y,\omega,x)G_\mathrm{me}^*(y,\omega,x')dy\nonumber\\
 &+k_0^2\int_{-\infty}^\infty\varepsilon(y,\omega)G_\mathrm{ee}(y,\omega,x)G_\mathrm{ee}^*(y,\omega,x')dy\nonumber\\
 &=-G_\mathrm{ee}^*(x,\omega,x').
\end{align}
By applying the Green's function reciprocity relations
$G_\mathrm{ee}(x,\omega,x')=G_\mathrm{ee}(x',\omega,x)$
and
$G_\mathrm{me}(x,\omega,x')=-G_\mathrm{em}(x',\omega,x)$,
we get
\begin{align}
 &-k_0^2\int_{-\infty}^\infty\mu^*(y,\omega)G_\mathrm{em}(x,\omega,y)G_\mathrm{em}^*(x',\omega,y)dy\nonumber\\
 &+k_0^2\int_{-\infty}^\infty\varepsilon(y,\omega)G_\mathrm{ee}(x,\omega,y)G_\mathrm{ee}^*(x',\omega,y)dy\nonumber\\
 &=-G_\mathrm{ee}^*(x,\omega,x').
\end{align}
Taking the imaginary part and switching the terms gives the final result
\begin{align}
 &k_0^2\int_{-\infty}^\infty\varepsilon_\mathrm{i}(y,\omega)G_\mathrm{ee}(x,\omega,y)G_\mathrm{ee}^*(x',\omega,y)dy\nonumber\\
 &+k_0^2\int_{-\infty}^\infty\mu_\mathrm{i}(y,\omega)G_\mathrm{em}(x,\omega,y)G_\mathrm{em}^*(x',\omega,y)dy\nonumber\\
 &=\mathrm{Im}[G_\mathrm{ee}(x,\omega,x')].
 \label{eq:greeneidentity}
\end{align}
Respectively, for the magnetic Green's function, we obtain
\begin{align}
 &k_0^2\int_{-\infty}^\infty\mu_\mathrm{i}(y,\omega)G_\mathrm{mm}(x,\omega,y)G_\mathrm{mm}^*(x',\omega,y)dy\nonumber\\
 &+k_0^2\int_{-\infty}^\infty\varepsilon_\mathrm{i}(y,\omega)G_\mathrm{me}(x,\omega,y)G_\mathrm{me}^*(x',\omega,y)dy\nonumber\\
 &=\mathrm{Im}[G_\mathrm{mm}(x,\omega,x')].
 \label{eq:greenmidentity}
\end{align}

\subsection{Integral identity for $k$ integration}
Here we derive the integral identity for $k$ integration used in one of the intermediate
steps in Eq.~\eqref{eq:commutatorderivation3}.
The integrand in
\begin{equation}
 \frac{1}{2\pi i}\int_{-\infty}^\infty\frac{e^{ik(x-x')}}{k^2-\omega^2 n^2/c^2}dk
\end{equation}
has two poles at positions $k=\pm\omega n/c$. The pole with positive sign
is located in the upper half of the complex $k$ plane and the pole with negative sign
is located in the lower half plane.
When $x>x'$ the integrand goes to zero in the upper half plane
as $k\rightarrow\infty$. Therefore, by using the residue theorem
we obtain
\begin{align}
 &\frac{1}{2\pi i}\int_{-\infty}^\infty\frac{e^{ik(x-x')}}{k^2-\omega^2 n^2/c^2}dk\nonumber\\
 &=\underset{k=\omega n/c}{\mathrm{Res}}\,
 \frac{e^{ik(x-x')}}{k^2-\omega^2 n^2/c^2}
 =\frac{e^{i\omega n(x-x')/c}}{2\omega n/c}.
 \label{eq:residue1}
\end{align}
When $x<x'$ the integrand, respectively, goes to zero in the lower half of the complex plane
as $k\rightarrow\infty$ and, by applying the residue theorem,
we obtain
\begin{align}
 &\frac{1}{2\pi i}\int_{-\infty}^\infty\frac{e^{ik(x-x')}}{k^2-\omega^2 n^2/c^2}dk\nonumber\\
 &=-\underset{k=-\omega n/c}{\mathrm{Res}}\,
 \frac{e^{ik(x-x')}}{k^2-\omega^2 n^2/c^2}
 =\frac{e^{-i\omega n(x-x')/c}}{2\omega n/c}.
 \label{eq:residue2}
\end{align}
The two equations in Eqs.~\eqref{eq:residue1} and \eqref{eq:residue2}
can be combined to give the final result
\begin{equation}
 \frac{1}{2\pi i}\int_{-\infty}^\infty\frac{e^{ik(x-x')}}{k^2-\omega^2 n^2/c^2}dk
 =\frac{e^{i\omega n|x-x'|/c}}{2\omega n/c}.
 \label{eq:math1}
\end{equation}

\subsection{Integral identity for $\omega$ integration}
Here we derive the integral identity for $\omega$ integration used in one of the intermediate
steps in Eq.~\eqref{eq:commutatorderivation3}.
The integrand in
\begin{equation}
 \int_{-\infty}^\infty\frac{\omega\mu(\omega)}{k^2c^2-\omega^2 n(\omega)^2}d\omega
 \label{eq:circle1}
\end{equation}
has no poles in the upper half of the complex $\omega$ plane.
The integral along the real $\omega$ axis in Eq.~\eqref{eq:circle1} is therefore
the negative of the integral around the semicircle at infinity
in the upper half plane, so putting $\omega=\Omega e^{i\varphi}$,
we obtain
\begin{align}
 &\int_{-\infty}^\infty\frac{\omega\mu(\omega)}{k^2c^2-\omega^2 n(\omega)^2}d\omega\nonumber\\
 &=-\lim_{\Omega\rightarrow\infty}\int_0^\pi\frac{i\mu(\Omega e^{2i\varphi})\Omega^2e^{2i\varphi}}{k^2c^2-\Omega^2n(\Omega e^{2i\varphi})^2e^{2i\varphi}}d\varphi\nonumber\\
 &=i\int_0^\pi d\varphi=i\pi
 \label{eq:circle2}
\end{align}
where we have applied the fact that material parameters are analytic
functions of frequency and become unity at high frequencies.
Thus, we have
\begin{equation}
 \int_{-\infty}^\infty\frac{\omega\mu(\omega)}{k^2c^2-\omega^2 n(\omega)^2}d\omega
 =i\pi.
 \label{eq:math2}
\end{equation}

\end{document}